\documentclass[12pt]{article}
\usepackage{epsfig}
\usepackage{amsmath}
\usepackage{indentfirst}
\usepackage{cite}
\usepackage{multirow}
\usepackage{amssymb}
\usepackage{titlesec}

\begin{document}

\renewcommand{\figurename}{\footnotesize{Figure}}
\renewcommand{\tablename}{\footnotesize{Table}}
\titleformat*{\section}{\fontsize{14}{17}\selectfont\bfseries}

\begin{center}
\textbf{\Large $\alpha$-cluster structure above double-shell closures and \\ $\alpha$-decay of $^{104}$Te}
\end{center}

\begin{center}
M.~A.~Souza$^{a,b,*}$, H.~Miyake$^{a}$, T.~Borello-Lewin$^{a}$, C.~A.~da Rocha$^{c}$, and C.~Frajuca$^{b}$

\

\begin{footnotesize}
$^{a}$ {\it Instituto de F\'{\i}sica, Universidade de S\~{a}o Paulo, Rua do Mat\~{a}o, 1371, CEP 05508-090, Cidade Universit\'{a}ria,
S\~{a}o Paulo - SP, Brazil} \\
\vspace{2pt}
$^{b}$ {\it Departamento de Mec\^{a}nica, Instituto Federal de Educa\c{c}\~{a}o, Ci\^{e}ncia e Tecnologia de S\~{a}o Paulo -
Campus S\~{a}o Paulo, Rua Pedro Vicente, 625, CEP 01109-010, Canind\'{e}, S\~{a}o Paulo - SP, Brazil} \\
\vspace{2pt}
$^{c}$ {\it Departamento de Ci\^{e}ncias e Matem\'{a}tica, Instituto Federal de Educa\c{c}\~{a}o, Ci\^{e}ncia e Tecnologia de S\~{a}o Paulo -
Campus S\~{a}o Paulo, Rua Pedro Vicente, 625, CEP 01109-010, Canind\'{e}, S\~{a}o Paulo - SP, Brazil} \\
\vspace{6pt}
* E-mail: marsouza@if.usp.br \\
\vspace{6pt}
\end{footnotesize}
\end{center}

\

\begin{quotation}
\begin{footnotesize}
\noindent {\bf Abstract:}

\noindent An analysis of the $\alpha$ + core properties of $^{104}$Te along with a global discussion on the $\alpha$-cluster structure above the double-shell closures is presented from the viewpoint of the local potential model. The $\alpha$ + core interaction is described by a nuclear potential of \linebreak[4] \mbox{(1 + Gaussian)$\times$(W.S.~+ W.S.$^3$)} shape with two free parameters, which has been successfully tested in nuclei of different mass regions. The model produces $Q_{\alpha}$ values and $\alpha$-decay half-lives for $^{104}$Te in agreement with the 2018 experimental data of Auranen {\it et al.}~in the energy range $5.13 \; \mathrm{MeV} < Q_{\alpha} < 5.3 \; \mathrm{MeV}$ using an $\alpha$ preformation factor $P_{\alpha} = 1$. The comparison of the calculated reduced $\alpha$-widths for the ground state bands of $^{104}$Te, $^{94}$Mo and $^{212}$Po indicates that $^{104}$Te has an $\alpha$-cluster degree significantly higher than $^{94}$Mo and much higher than $^{212}$Po. These results point to $^{104}$Te as a preferential nucleus for $\alpha$-clustering in the $N,Z = 50$ region and corroborate the statement on the superallowed $\alpha$-decay in $^{104}$Te.
\end{footnotesize}
\end{quotation}

\

\section{Introduction}

The $\alpha$-cluster model has been used to successfully describe the spectroscopic properties of nuclei from different mass regions, with different approaches, consolidating itself as an important topic of the nuclear structure. The two reviews in Refs.~\cite{FHK2018,RZ2018} show the current applications and methods in this research area. The binary \mbox{$\alpha$ + core} structure has been studied in several works, with emphasis on nuclei in which the core has doubly closed shell, such as $^{20}$Ne, $^{44}$Ti, $^{60}$Zn, $^{94}$Mo, and $^{212}$Po (examples in Refs. \cite{BMP95,MOR1998,BJM1995,BMP1999_PRC61,MRO2000,SM2015,HMS1994,KWI2018}). In this context, there is a great interest in the $^{104}$Te nucleus, since an $\alpha$-decay experimental observation of this nucleus can indicate the presence of the $\alpha + ^{100}$Sn structure, as well as provide information on the nuclear structure of this mass region due to the double shell closure $N,Z = 50$. Experimental works such as Liddick {\it et al.}~\cite{LGM2006}, Janas {\it et al.}~\cite{JMB2005}, and Auranen {\it et al.}~\cite{ASA2018}, present experimental results of $\alpha$-decay chains in the vicinity of the doubly-magic nucleus $^{100}$Sn. Such decay processes are interpreted as superallowed $\alpha$-decays, since they indicate much larger $\alpha$-particle preformation factors than those seen in $\alpha$-decays in the vicinity of $^{208}$Pb. 

The observation of the $^{108}\mathrm{Xe} \rightarrow$ $^{104}\mathrm{Te} \rightarrow$ $^{100}\mathrm{Sn}$ $\alpha$-decay chain was recently reported \cite{ASA2018}, including the measurement of the $\alpha$-decay energy and half-life of $^{104}$Te ($Q_{\alpha}^{\mathrm{exp}} = 5.1(2)$ MeV and $T_{1/2}^{\mathrm{exp}} < 18$ ns). In the same reference, a comparative study shows that the $\alpha$-particle preformation factor for $^{108}$Xe or $^{104}$Te is more than a factor of 5 larger than that for $^{212}$Po, indicating superallowed $\alpha$-decay for those nuclei. The experimental information of Ref.~\cite{ASA2018} motivates the comparison with nuclear models.

Calculations on $^{104}$Te were made before the measurement of Ref.~\cite{ASA2018}. E.g., a microscopic description of $\alpha$-decay transitions in Te isotopes \cite{PLW2016} points out that the $\alpha$-particle formation probability for $^{104}$Te is a few times larger than for $^{212}$Po, reinforcing the interpretation of superallowed $\alpha$-decay for $^{104}$Te. The 2007 work of P.~Mohr \cite{M2007} discusses the $\alpha$-decay in Te and Xe isotopes and shows predictions on the \mbox{$\alpha$ + core} structure in $^{104}$Te by using a double-folding potential for the nuclear $\alpha$ + core interaction. Mohr obtained the theoretical energy $Q_{\alpha} = 5.42 \pm 0.07$ MeV and the $\alpha$-decay half-life $T_{1/2,\alpha} \approx 5$ ns with an $\alpha$ preformation factor $P_{\alpha} = 10 \; \%$ (with an uncertainty of a factor of two for $P_{\alpha}$). The $T_{1/2,\alpha}$ value predicted by Mohr is quantitatively in agreement with the experimental measure of Ref.~\cite{ASA2018}, however the predicted value for $Q_{\alpha}$ is a little above that obtained in the recent experimental measurement.

The recent work of D.~Bai and Z.~Ren \cite{BR2018} shows an analysis of the $\alpha$-cluster structure in $^{104}$Te and $^{108}$Xe in the framework of the density-dependent cluster model, taking into account the experimental results of Ref.~\cite{ASA2018}. The authors obtain the half-life $T_{1/2} (^{104}\mathrm{Te}) =$ 7$-$166 ns corresponding to the experimental range of $Q_{\alpha}$, and $T_{1/2}(^{104}\mathrm{Te}) = 32$ ns for the central experimental energy $Q_{\alpha}(^{104}\mathrm{Te}) = 5.1$ MeV assuming an $\alpha$ preformation factor $P_{\alpha} = 1$. However, the $\alpha$-branching ratios $b_{\alpha}$ obtained for the ground state band suffer strong variations with the change of intensity of the $L$-dependent renormalization factor of the double-folding potential.

With the new experimental data of Auranen {\it et al.}, a more comprehensive discussion of the $\alpha$-cluster structure above double-shell closures can be made. In the present work, the $\alpha$ + core structure in $^{104}$Te is analyzed with a nuclear potential of \mbox{(1 + Gaussian)$\times$(W.S.~+ W.S.$^3$)} shape, along with a global discussion on the $\alpha$-cluster structure in $^{20}$Ne, $^{44}$Ti, $^{94}$Mo, and $^{212}$Po.

\section{$\alpha $-cluster model}

The $\alpha$-cluster model assumes the total nucleus as an $\alpha$-particle orbiting an inert core. The $\alpha$ + core interaction is described through the local potential \mbox{$V(r)=V_N(r)+V_C(r)$} containing the nuclear and Coulomb terms. The Coulomb potential $V_C(r)$ is that of an $\alpha$-particle interacting with an uniformly charged spherical core of radius $R$. The intercluster nuclear potential $V_N(r)$ is expressed by

\begin{eqnarray}
V_N(r) = -V_{0}\left[1+\lambda\exp\left(-\frac{r^{2}}{\sigma^{2}}\right)\right]
 \left\{ \frac b{1+\exp [(r-R)/a]} \right. {}
                                                          \nonumber\\[4pt]
 {} \left. + \frac{1-b}{\{1+\exp[(r-R)/3a]\}^3}\right\} \;, \qquad \qquad
\label{eq:Nuc_Pot}
\end{eqnarray}

\noindent where $V_0$, $\lambda$, $a$, and $b$ are fixed parameters, and $R$ and $\sigma$ are variable parameters. This potential shape, identified here by \mbox{(1 + Gaussian)$\times$(W.S.~+ W.S.$^3$)}, was first proposed by Souza and Miyake for the analysis of the $\alpha$-cluster structure in $^{46}$Cr and $^{54}$Cr \cite{SM2017}, obtaining a good description of the ground state bands and $B(E2)$ transition rates for the two nuclei.

The \mbox{(1 + Gaussian)$\times$(W.S.~+ W.S.$^3$)} potential was developed from the \mbox{W.S.~+ W.S.$^3$} nuclear potential previously used by Buck, Merchant, and Perez for the investigation of the $\alpha$-cluster structure in $^{20}$Ne, $^{44}$Ti, $^{94}$Mo, and $^{212}$Po \cite{BMP95}. The calculation of Ref.~\cite{BMP95}, although successful in the general description of the ground state bands of the mentioned nuclei, roughly reproduces the energies of the $0^{+}$ bandheads. It is to be noted that the inclusion of the proposed \mbox{(1 + Gaussian)} factor in the \mbox{W.S.~+ W.S.$^3$} nuclear potential, while relevant for the correct description of the $0^{+}$ bandheads, produces very minor changes in the energy levels above $0^{+}$ (e.g., see Fig.~3 of Ref.~\cite{SM2017}).

The parameter values used are: $V_0 = 220$ MeV, $a = 0.65$ fm, $b = 0.3$, and $\lambda = 0.14$, while $R$ and $\sigma$ are fitted specifically for each nucleus. The values of $V_0$, $a$ and $b$ are the same used in Refs.~\cite{SM2015,BMP95} to describe the ground state bands of nuclei of different mass regions with the W.S.$+$W.S.$^{3}$ nuclear potential. The parameter $\lambda = 0.14$ applied is a mean value suitable for reproducing the $0^{+}$ bandheads of the ground state bands of $^{20}$Ne, $^{44}$Ti, $^{94}$Mo, and $^{212}$Po, taking the corresponding $R$ values obtained from Ref.~\cite{SM2015}. The variable parameters $\sigma$ and $R$, shown in Table \ref{Table_parameters}, are fitted to reproduce the experimental 0$^{+}$ and 4$^{+}$ members of the ground state band.

\begin{table}
\caption{Values of the parameters $R$ and $\sigma$ for the $^{20}$Ne, $^{44}$Ti, $^{94}$Mo, and $^{212}$Po nuclei.}
\label{Table_parameters}
\begin{center}
\begin{tabular}{ccc}
\hline
&  &  \\[-12pt]
Nucleus & $R$ (fm) & $\sigma$ (fm) \\[2pt]
\hline
&  &  \\[-10pt] 
$^{20}$Ne & 3.272 & 0.475 \\
$^{44}$Ti & 4.551 & 0.425 \\
$^{94}$Mo & 5.783 & 0.410 \\
$^{212}$Po & 7.018 & 0.445 \\
\hline
\end{tabular}
\end{center}
\end{table}

\begin{figure}[t]
\centering
\includegraphics[scale=0.64]{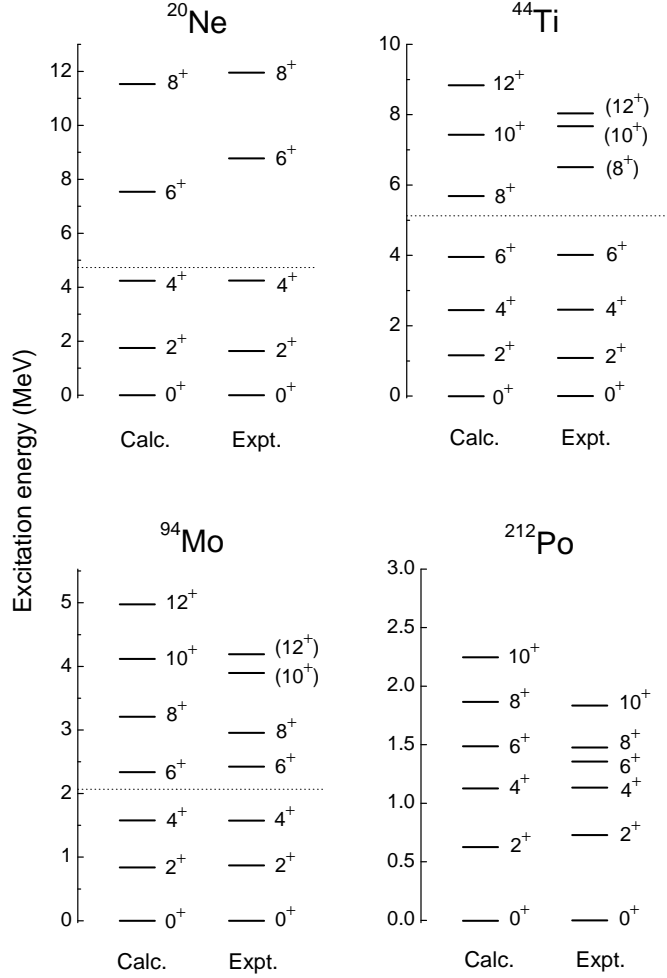}
\caption{Calculated energy levels for the ground state bands of $^{20}$Ne ($G = 8$), $^{44}$Ti ($G = 12$), $^{94}$Mo ($G = 16$), and $^{212}$Po ($G = 20$) in comparison with experimental energies. The dotted lines indicate the $\alpha$ + core thresholds.}
\label{Fig_spectra}
\end{figure}

The nucleons of the $\alpha$-cluster must lie in shell-model orbitals outside the core. This restriction is defined by the global quantum number $G = 2N + L$, where $N$ is the number of internal nodes in the radial wave function and $L$ is the orbital angular momentum. The numbers $G = 8$, 12, and 16 are applied to the ground state bands of $^{20}$Ne, $^{44}$Ti, and $^{94}$Mo, respectively, as obtained by the Wildermuth condition \cite{WT1977}. $G = 20$ is applied to the $^{212}$Po ground state band, since this number is cited in Ref.~\cite{BMP95} as the more appropriate to describe the experimental spectrum of $^{212}$Po in comparison with the quantum number predicted by the Wildermuth condition.

\section{Results and discussion}

Figure \ref{Fig_spectra} shows the ground state bands calculated for $^{20}$Ne, $^{44}$Ti, $^{94}$Mo, and $^{212}$Po compared to the experimental energies. The \mbox{(1 + Gaussian)$\times$(W.S.~+ W.S.$^3$)} potential maintains a good general description of the levels above $0^{+}$ and additionally allows the correct reproduction of the $0^{+}$ bandheads. Also, the \mbox{(1 + Gaussian)$\times$(W.S.~+ W.S.$^3$)} potential has the merit of satisfactorily reproducing the spectra of the four nuclei without the inclusion of a $L$-dependent strength parameter.

The $\alpha$ + core model applied to $^{104}$Te resulted in the properties shown in Table \ref{Table_widths}. The energy levels of the ground state band are calculated with the variable parameters: $\sigma = 0.44$ fm and $R = 5.713$ fm. As only the ground state $0^{+}$ has been measured experimentally, the value of $\sigma$ for $^{104}$Te is obtained from the mean value for the same parameter in the set \{$^{20}$Ne, $^{44}$Ti, $^{94}$Mo, $^{212}$Po\}. The value of $R$ was fitted to reproduce the energy $E(0^{+}) = 5.22$ MeV, which is compatible with the experimental energy $Q_{\alpha} = 5.1(2)$ MeV. The energy $E(0^{+}) = 5.22$ MeV was chosen because it is between the minimum admissible theoretical energy for the $0^{+}$ state (5.13 MeV, where $T_{1/2,\alpha} \approx 18$ ns) and the maximum energy resulting from the experimental uncertainty (5.3 MeV). The quantum number $G = 16$ was used for the ground state band, according to the Wildermuth condition considering the $(sdg)^4$ configuration.

\begin{table}
\caption{Calculated energy levels ($E$), rms intercluster separations ($\langle R^2 \rangle ^{1/2}$), reduced $\alpha$-widths ($\gamma _\alpha ^2$), $B(E2)$ transition rates, $\gamma$-decay widths ($\Gamma_{\gamma}$), $\alpha$-decay widths ($\Gamma_{\alpha}$), and $\alpha$-branching ratios ($b_{\alpha}$) for the ground state band of $^{104}$Te. The values adopted for the internal conversion coefficient ($\alpha_{IC}$) are indicated. An $\alpha$ preformation factor $P_{\alpha} = 1$ is applied. The energy levels are given with reference to the $\alpha + ^{100}$Sn threshold.}
\label{Table_widths}
\begin{footnotesize}
\begin{center}
\begin{tabular}{ccccccccc}
\hline
&  &  &  &  &  &  &  &  \\[-10pt]
       & $E$ & $\langle R^2 \rangle ^{1/2}$ & $\gamma _\alpha ^2$ & $B(E2;J\rightarrow J-2)$ & $\alpha_{IC}$ &  $\Gamma_{\gamma}(J\rightarrow J-2)$  &	$\Gamma_{\alpha}$ & $b_{\alpha}$ \\[2pt]
$J^\pi$ & (MeV)  &  (fm)  &  (eV)  &  ($e^2 \,$fm$^4$)  &  (Ref.~\cite{KBT2008})  &  (MeV)  &  (MeV)  & (\%) \\[2pt]
\hline
&  &  &  &  &  &  &  &  \\[-10pt]
$0^{+}$	  &  5.220	&  5.249 &  1009.6 &			 &	          &              &  5.146E$-14$ & 100.00 \\
$2^{+}$	  &  6.046	&  5.262 &  1130.4 &   224.23	 &  0.002234  &	 6.968E$-11$ &  6.861E$-12$ &   8.96  \\
$4^{+}$	  &  6.889	&  5.232 &  953.3  &   312.10	 &  0.00213	  &  1.074E$-10$ &  1.206E$-10$ &  52.90  \\
$6^{+}$	  &  7.758	&  5.156 &  598.3  &   313.88   &	0.001985  &  1.257E$-10$ &  4.357E$-10$ &  77.61  \\
$8^{+}$	  &  8.757	&  5.055 &  287.5  &   283.03   &	0.00145	  &  2.274E$-10$ &  6.231E$-10$ &  73.26  \\
$10^{+}$  &  9.820	&  4.943 &  100.7  &   234.88   &	0.001266  &  2.574E$-10$ &  3.159E$-10$ &  55.10  \\
$12^{+}$  &  10.833	&  4.834 &  23.8   &   177.83   &	0.001406  &  1.532E$-10$ &  4.541E$-11$ &  22.87  \\
$14^{+}$  &  11.643	&  4.745 &  3.3    &   117.56   &	0.002339  &  3.313E$-11$ &  1.399E$-12$ &   4.05  \\
$16^{+}$  &  12.033	&  4.696 &  0.2    &    57.63   &	0.01691	  &  4.264E$-13$ &  4.575E$-15$ &   1.06  \\
\hline
\end{tabular}
\end{center}
\end{footnotesize}
\end{table}

The $B(E2)$ transition rates are calculated from the radial wave functions $u(r)$ of the $\alpha$ + core states, according to the formulae presented in Ref.~\cite{SM2015}. In its turn, the $B(E2)$ values are used for calculating the $\gamma$-decay widths ($\Gamma_{\gamma}$), according to the formula presented in Ref.~\cite{HMS1994}. The $\alpha$-decay widths ($\Gamma_{\alpha}$) are calculated by the semi-classical approximation of Ref.~\cite{GK1987}, using an $\alpha$ preformation factor $P_{\alpha} = 1$. The $\alpha$-branching ratio $b_{\alpha}$ is given by \mbox{$b_{\alpha} = \Gamma_{\alpha}/(\Gamma_{\alpha} + \Gamma_{\gamma})$}.

The internal conversion process is taken into account through the internal conversion coefficient $\alpha_{IC}$. For low transition energies, the internal conversion coefficient can exert a major influence on the value of $\Gamma_{\gamma}$, which is proportional to the factor \mbox{($1 + \alpha_{IC}$)} \cite{HMS1994}. In this work, the values of $\alpha_{IC}$ are obtained from the internal conversion coefficient database $BrIcc$ \cite{KBT2008}. In the case of $^{104}$Te, the theoretical transition energies \mbox{$E_{\gamma} = E(J) - E(J-2)$} are large enough so that the influence of the internal conversion process is very small, as seen by the small values of $\alpha_{IC}$ shown in Table \ref{Table_widths}.

The $\alpha$-decay half-life, given by $T_{1/2,\alpha} = \hbar \ln 2 / \Gamma_{\alpha}$, is 8.866 ns for $E(0^{+}) = 5.22$ MeV; this $T_{1/2,\alpha}$ value is compatible with the experimental measure of Ref.~\cite{ASA2018}. By varying only the parameter $R$, it is verified that the energy range $5.13 \; \mathrm{MeV} < E(0^{+}) < 5.3$ MeV provides $T_{1/2,\alpha}$ values compatible with the experimental measure, and simultaneously, is compatible with the experimental measure of $Q_{\alpha}$. The production of such results with $P_{\alpha} = 1$ corroborates the indication of the superallowed $\alpha$-decay of $^{104}$Te by Auranen {\it et al.}~\cite{ASA2018}. In addition, it should be noted that the correspondence with the experimental data is obtained with an $\alpha$ + core potential which showed to be successful in nuclei of other mass regions with the same fixed parameters and without $L$-dependent parameters.

\begin{figure}
\centering
\includegraphics[scale=0.7]{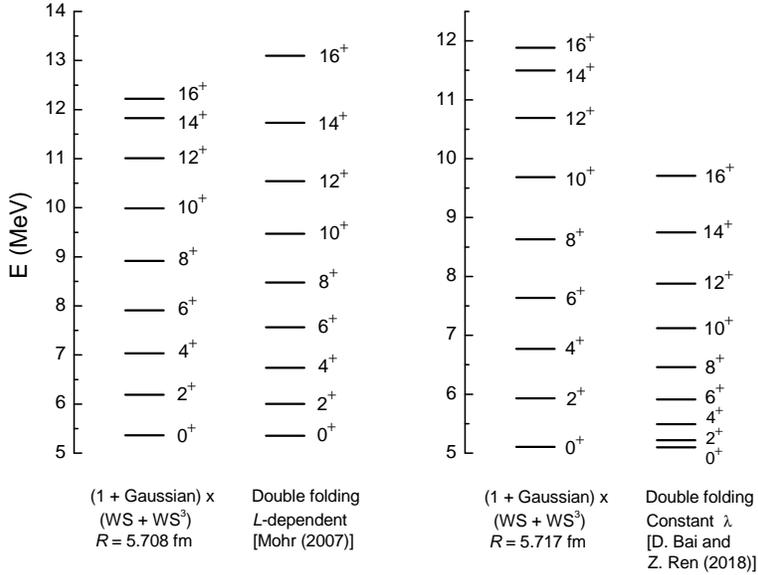}
\caption{Calculated energy levels for the ground state band ($G = 16$) of the $\alpha + ^{100}$Sn system in comparison with the theoretical bands calculated by P.~Mohr \cite{M2007} and D.~Bai and Z.~Ren \cite{BR2018} which use the double-folding potential as the nuclear $\alpha$ + core interaction. The calculation of Mohr includes a $L$-dependent strength parameter, and the calculation of D.~Bai and Z.~Ren applies a fixed strength parameter. The energy scale is given with reference to the $\alpha + ^{100}$Sn threshold.}
\label{Fig_energies_104Te}
\end{figure}

The values obtained for the $\alpha$-branching ratios indicate a predominance of the $\gamma$-decay in the first excited state ($2^{+}$). However, in the states from $4^{+}$ to $10^{+}$, there is a predominance of the $\alpha$-decay. This result indicates that the production of $^{104}$Te in its ground state is favored by the $2^{+} \rightarrow 0^{+}$ transition. It is to be emphasized that the values of $\Gamma_{\gamma}$, $\Gamma_{\alpha}$, and $b_{\alpha}$ in Table \ref{Table_widths} are calculated in accordance with the experimental data of $Q_{\alpha}$ and $T_{1/2}$ for $^{104}$Te. The results of Table \ref{Table_widths} differ significantly from the results presented by D.~Bai and Z.~Ren for the same system \cite{BR2018} using the double-folding potential for the $\alpha$ + core interaction and a fixed strength parameter. In Ref.~\cite{BR2018}, the energy levels produced with fixed strength parameter are more compressed and have a quasi-rotational behavior (see Fig.~\ref{Fig_energies_104Te}), contributing to a difference of orders of magnitude for the $\Gamma_{\gamma}$ and $\Gamma_{\alpha}$ values of Ref.~\cite{BR2018} and the present work. The $b_{\alpha}$ values with fixed strength parameter of Ref.~\cite{BR2018} also differ considerably from the respective values of this work, indicating the predominance of the $\alpha$-decay in the $2^{+}$ state and a very high predominance of the $\gamma$-decay for the states above $2^{+}$. However, the inclusion of an $L$-dependent strength parameter in the double-folding potential produces $b_{\alpha}$ values a little closer to these of the present work at the $2^{+}$ and $4^{+}$ states, even changing the predominance of the $2^{+}$ state to the $\gamma$-decay.

Figure \ref{Fig_energies_104Te} shows the comparison between the spectra produced by the (1 + Gaussian)$\times$\linebreak[4](W.S.~+ W.S.$^3$) and double-folding potentials for the ground state band. The first comparison is made with the calculation of Mohr \cite{M2007} with double-folding potential and $L$-dependent strength parameter, and the second comparison is made with the calculation of D.~Bai and Z.~Ren \cite{BR2018} with double-folding potential and fixed strength parameter. In the two comparisons, the parameter $R$ of the \mbox{(1 + Gaussian)$\times$(W.S.~+ W.S.$^3$)} potential is smoothly changed to reproduce the $E(0^{+})$ energies suggested in the previous calculations. On the comparison with the calculation of Mohr, a similarity is observed between the two theoretical spectra, except for the spacing between the $14^{+}$ and $16^{+}$ levels. On the comparison with the calculation of D.~Bai and Z.~Ren, there is a significant difference between the two theoretical spectra since the double-folding potential with fixed strength parameter produces a more compressed spectrum with the rotational feature. In previous works \cite{MOR1998,O1995,M2017}, it was shown that the double-folding potential requires a $L$-dependent strength parameter for the satisfactory reproduction of the experimental spectra of different nuclei.

The calculated rms intercluster separations ($\langle R^2 \rangle ^{1/2}$) and reduced $\alpha$-widths ($\gamma _\alpha ^2$) for the $^{104}$Te ground state band are shown in Table \ref{Table_widths}. The reduced $\alpha $-width is defined in the context of the $R$-matrix theory \cite{AY1974,MSV1970},

\begin{equation}
\gamma _{\alpha}^2=\left( \frac{\hbar^2}{2\mu a_{c}}\right) u^2(a_c)\left[
\int_0^{a_c}|u(r)|^2dr\right] ^{-1}\;,
\label{Red_width}
\end{equation}

\noindent where $\mu $ is the reduced mass of the system, $u(r)$ is the radial wave function of the state, and $a_c$ is the channel radius, which is obtained from a linear fit that considers other channel radii of different $\alpha$ + core systems in the literature (see eq.~(12) of Ref.~\cite{SM2015}). The $\langle R^2 \rangle ^{1/2}$ and $\gamma _\alpha ^2$ values indicate a stronger $\alpha$-cluster character for the first members of the band and the well-known antistretching effect, which is observed with other $\alpha$ + core potential forms \cite{SM2015,MOR1998,O1995,M2017}. It is important to note that the inclusion of the factor \mbox{(1 + Gaussian)} in the \mbox{W.S.~+ W.S.$^3$} potential does not change this general property.

Table \ref{Table_94Mo_212Po} shows a comparison of the reduced $\alpha$-widths obtained for the ground state bands of $^{104}$Te, $^{94}$Mo and $^{212}$Po through the ratios \mbox{$\gamma _\alpha ^2$($^{104}$Te)/$\gamma _\alpha ^2$($^{94}$Mo)} and \mbox{$\gamma _\alpha ^2$($^{104}$Te)/$\gamma _\alpha ^2$($^{212}$Po)}. Such a comparison is useful since $^{94}$Mo and $^{212}$Po are regarded as preferential nuclei for the $\alpha$ + core structure in their respective mass regions because the $\alpha$-clustering occurs above the double-shell closures at $^{90}$Zr and $^{208}$Pb, respectively. The reduced $\alpha$-widths of $^{94}$Mo and $^{212}$Po are calculated with eq.~\eqref{Red_width} and channel radii given by eq.~(12) of Ref.~\cite{SM2015}. For the calculation of the radial wave functions $u(r)$, the parameter $V_0$ is smoothy modified to reproduce experimental energy levels of $^{94}$Mo and $^{212}$Po from $0^{+}$ to $10^{+}$. The ratios \mbox{$\gamma _\alpha ^2$($^{104}$Te)/$\gamma _\alpha ^2$($^{94}$Mo)} vary around 1.5, and the ratios \mbox{$\gamma _\alpha ^2$($^{104}$Te)/$\gamma _\alpha ^2$($^{212}$Po)} vary in the range from $\approx$ 17.7 to $\approx$ 20.5. In the case of \mbox{$\gamma _\alpha ^2$($^{104}$Te)/$\gamma _\alpha ^2$($^{94}$Mo)}, it is suggested that $^{104}$Te has a considerably higher degree of $\alpha$-clustering in comparison with $^{94}$Mo; this is important information on the $\alpha$-cluster structure in the $N,Z = 50$ region, since $^{94}$Mo is regarded in previous works as a preferential nucleus for $\alpha$-clustering in this region. The ratios \mbox{$\gamma _\alpha ^2$($^{104}$Te)/$\gamma _\alpha ^2$($^{212}$Po)} indicate a much higher degree of $\alpha$-clus\-ter\-ing for $^{104}$Te than for $^{212}$Po. Auranen {\it et al.}~\cite{ASA2018} make a similar comparison of $^{104}$Te and $^{212}$Po using the definition of reduced $\alpha$-decay width (identified as $\delta^2$) from Ref.~\cite{R1959}. In Ref.~\cite{ASA2018}, the ratio \mbox{$\delta^2$($^{104}$Te)/$\delta^2$($^{212}$Po)} $\gtrsim 13.1$, showing an agreement with the prediction of this work. Therefore, the ratios \mbox{$\gamma _\alpha ^2$($^{104}$Te)/$\gamma _\alpha ^2$($^{212}$Po)} suggest once again the superallowed $\alpha$-decay for $^{104}$Te.

\begin{table}
\caption{Calculated reduced $\alpha $-widths ($\gamma _\alpha ^2$) for the ground state bands of $^{94}$Mo ($G = 16$) and $^{212}$Po ($G = 20$) from the $0^{+}$ state to $10^{+}$ state, and ratios \mbox{$\gamma _\alpha ^2$($^{104}$Te)/$\gamma _\alpha ^2$($^{94}$Mo)} and \mbox{$\gamma _\alpha ^2$($^{104}$Te)/$\gamma _\alpha ^2$($^{212}$Po)} for the corresponding states.}
\label{Table_94Mo_212Po}
\begin{center}
\begin{tabular}{ccccc}
\hline
&  &  &  &  \\[-12pt]
	 & $\gamma _\alpha ^2$($^{94}$Mo) & $\gamma _\alpha ^2$($^{212}$Po) & \multirow{2}{*}{$\displaystyle{\frac{\gamma_\alpha ^2(^{104}\mathrm{Te})}{\gamma_\alpha ^2(^{94}\mathrm{Mo})}}$} & \multirow{2}{*}{$\displaystyle{\frac{\gamma_\alpha ^2(^{104}\mathrm{Te})}{\gamma_\alpha ^2(^{212}\mathrm{Po})}}$} \\
$J^\pi $  &  (eV)  &	   (eV)   &	  &	  \\[4pt]
\hline
&  &  &  &  \\[-10pt]
$0^{+}$   & 685.8  &  55.198  &	1.472  &  18.291 \\
$2^{+}$   & 757.8  &  63.851  &	1.492  &  17.704 \\
$4^{+}$   & 613.4  &  51.464  &	1.554  &  18.524 \\
$6^{+}$   & 397.2  &  30.615  &	1.506  &  19.543 \\
$8^{+}$   & 177.4  &  14.011  &	1.621  &  20.520 \\
$10^{+}$  & 65.9   &  5.561   &	1.528  &  18.108 \\
\hline
\end{tabular}
\end{center}
\end{table}

\begin{figure}[t]
\centering
\includegraphics[scale=0.58]{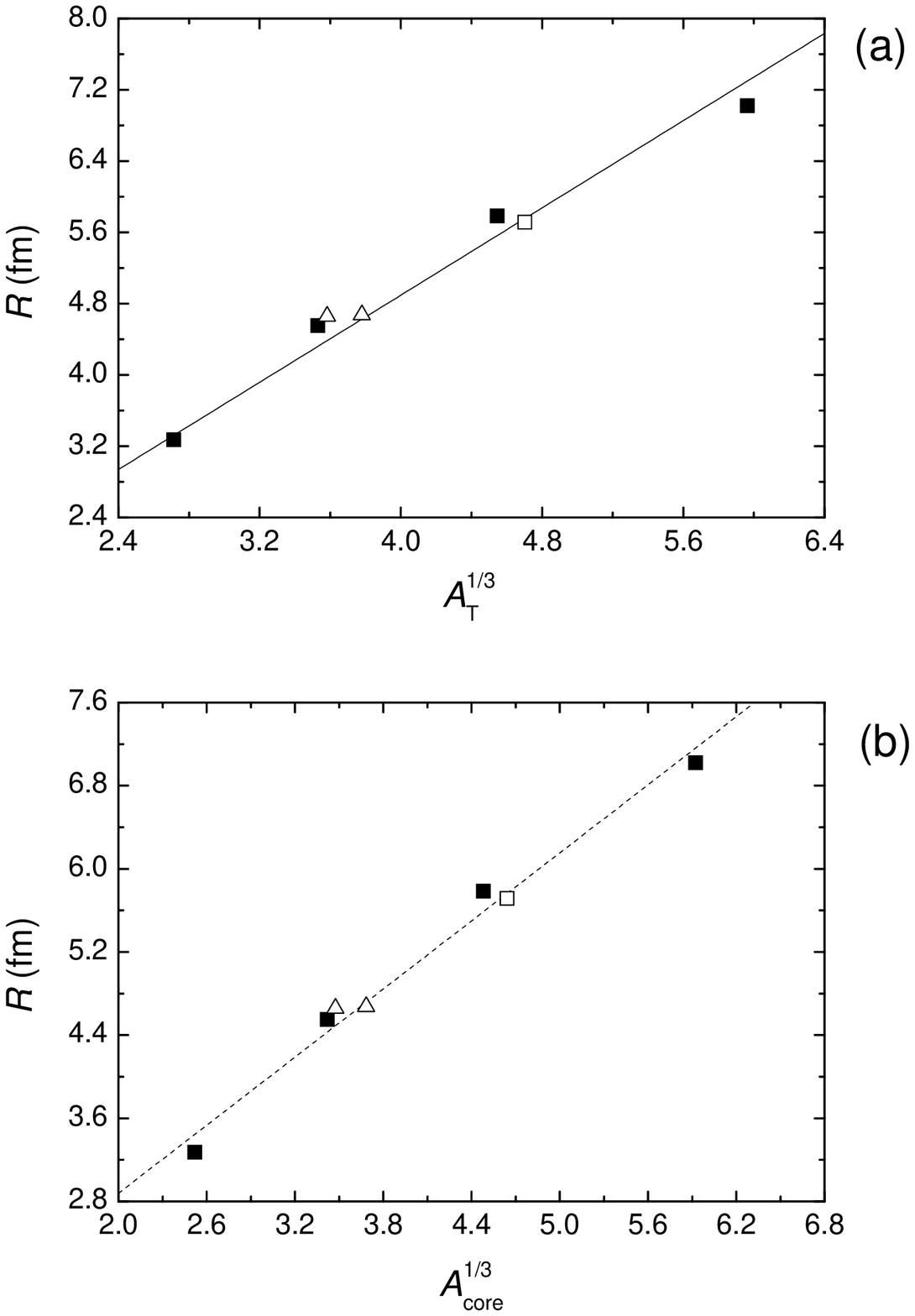}
\caption{Values of the parameter $R$ as a function of $A_{T}^{1/3}$ {\bf (a)} and $A_{\mathrm{core}}^{1/3}$ {\bf (b)}. The full squares correspond to the set \{$^{20}$Ne, $^{44}$Ti, $^{94}$Mo, $^{212}$Po\}; the open square corresponds to the value $R = 5.713$ fm of $^{104}$Te; the open triangles correspond to the $^{46}$Cr and $^{54}$Cr nuclei which were analyzed with the \mbox{(1 + Gaussian)$\times$(W.S.~+ W.S.$^3$)} potential in Ref.~\cite{SM2017}. The full line corresponds to the linear fit given by \mbox{$R = 1.224 \, A_{T}^{1/3}$} (fm), and the short dashed line corresponds to the linear fit given by \mbox{$R = 0.694 + 1.092 \, A_{\mathrm{core}}^{1/3}$} (fm).}
\label{Fig_R_fits}
\end{figure}

Next, an analysis of the radial parameter $R$ is presented. The values of $R$ are shown graphically in Fig.~\ref{Fig_R_fits} as a function of $A_{T}^{1/3}$ (graph {\bf (a)}) and $A_{\mathrm{core}}^{1/3}$ (graph {\bf (b)}), where $A_{T}^{1/3}$ and $A_{\mathrm{core}}^{1/3}$ are the mass numbers of the total nucleus and the core, respectively. Linear fits were made for the points referring to the set \{$^{20}$Ne, $^{44}$Ti, $^{94}$Mo, $^{212}$Po\}, with 1 or 2 adjustable parameters. The fits with the smallest standard deviations are shown graphically in Fig.~\ref{Fig_R_fits}. It can be seen that the two fitted functions provide a satisfactory description of the variation of $R$. The value of $R$ for $^{104}$Te is very close to the fitted functions, demonstrating a coherence of the applied value ($R = 5.713 $ fm) with the linear trend of $R$ through the different mass regions. The $R$ values for $^{46}$Cr and $^{54}$Cr, obtained from the previous work of Souza and Miyake \cite{SM2017} with use of the same $\alpha$ + core potential, are also close to the fitted lines. An important finding on the fits of Fig.~\ref{Fig_R_fits} is that the inclusion of the \mbox{(1 + Gaussian)} factor in the nuclear potential does not change the linear trend of $R$ in relation to the total nucleus and core radii, as previously observed with the simple \mbox{W.S.~+ W.S.$^3$} potential in Ref.~\cite{SM2015}.

\section{Summary and prospects}

It is shown that the \mbox{(1 + Gaussian)$\times$(W.S.~+ W.S.$^3$)} nuclear potential provides a good general description of the ground state bands of $^{20}$Ne, $^{44}$Ti, $^{94}$Mo, and $^{212}$Po with four fixed and two variable parameters. This potential shape allows the correct reproduction of the $0^{+}$ bandheads, representing an improvement over the \mbox{W.S.~+ W.S.$^3$} potential. The \mbox{(1 + Gaussian)$\times$(W.S.~+ W.S.$^3$)} potential produces solutions ($Q_{\alpha}, T_{1/2,\alpha}$) compatible with the experimental data in the energy range $5.13 \; \mathrm{MeV} < E(0^{+}) < 5.3 \; \mathrm{MeV}$ with \mbox{$P_{\alpha} = 1$}. The indication that $P_{\alpha} \sim 1$ corroborates the statement of Auranen {\it et al.}~\cite{ASA2018} on the superallowed $\alpha$-decay in $^{104}$Te. The study on the radial parameter $R$ shows that it has a linear trend relative to $A_T^{1/3}$ and $A_{\mathrm{core}}^{1/3}$, even with the inclusion of the factor \mbox{(1 + Gaussian)} on the \mbox{W.S.~+ W.S.$^3$} potential. The value $R = 5.713$ fm applied to $^{104}$Te is consistent with such linear trend. The calculated $\alpha$-branching ratios suggest that the production of $^{104}$Te in its ground state is favored by the $\gamma$-decay from the $2^{+}$ state. The ratios \mbox{$\gamma _\alpha ^2$($^{104}$Te)/$\gamma _\alpha ^2$($^{94}$Mo)} suggest that $^{104}$Te has an $\alpha$-cluster degree significantly higher than $^{94}$Mo, an important result concerning the $\alpha$-cluster structure in the $N,Z = 50$ region. The ratios \mbox{$\gamma _\alpha ^2$($^{104}$Te)/$\gamma _\alpha ^2$($^{212}$Po)} indicate that the $\alpha$-cluster degree in $^{104}$Te is much more pronounced than in $^{212}$Po, in agreement with deductions of Ref. \cite{ASA2018}. Such a result corroborates once again the statement on the superallowed $\alpha$-decay in $^{104}$Te.

There is a strong indication that the \mbox{(1 + Gaussian)$\times$(W.S.~+ W.S.$^3$)} potential can be applied glob\-al\-ly to describe the $\alpha$ + core structure in nuclei of distinct mass regions. The previous study on $^{46,54}$Cr \cite{SM2017} is a first demonstration that this potential works well in nuclei farther from the double-shell closures. Additional work as \cite{SMB2018} is in development to test its comprehensiveness.

\section*{Acknowledgments}
M.~A.~Souza receives financial support from Coordena\c{c}\~{a}o de A\-per\-fei\-\c{c}o\-a\-men\-to de Pessoal de N\'{i}vel Superior (CAPES). C.~Frajuca acknowledges Fun\-da\-\c{c}\~{a}o de Am\-pa\-ro \`{a} Pes\-qui\-sa do Es\-ta\-do de S\~{a}o Paulo (FAPESP) for grant \mbox{\#2013/26258-4} and Conselho Nacional de Desenvolvimento Cient\'{\i}fico e Tecnol\'{o}gico (CNPq) for grant \mbox{\#309098/2017-3}. Support from Instituto Nacional de Ci\^{e}ncia e Tecnologia - F\'{\i}sica Nuclear e Aplica\c{c}\~{o}es (INCT-FNA), grant \mbox{464898/2014-5}, is acknowledged. Research developed with HPC resources provided by Superintend\^{e}ncia de Tecnologia da Informa\c{c}\~{a}o da Universidade de S\~{a}o Paulo.


\end{document}